\documentclass[12pt,a4paper]{article}
\usepackage{ifpdf}
\usepackage{times}
\usepackage[scale={0.8,0.9},centering,includeheadfoot]{geometry}
\usepackage{enumitem}
\usepackage{subcaption}
\usepackage{multicol}
\usepackage{lscape}
\usepackage{listings}
\usepackage{float}
\usepackage{booktabs}
\usepackage{url}
\usepackage{graphicx}
\usepackage{multirow}

\usepackage{pgf}

\usepackage{amsmath,amssymb,xspace,amsthm,bm}
\usepackage{hyperref}

\usepackage[colon,longnamesfirst]{natbib}

\usepackage{dcolumn}\newcolumntype{d}[1]{D{.}{.}{#1}}

\newtheorem{proposition}{Proposition}
\newtheorem{remark}{Remark}

\def\qedbox{\ifvmode\else\unskip\fi~\penalty10000
    \hfill{\large$\blacksquare$}}

\def\miscinfo#1#2{{\footnotesize\indent\textsc{#1: }\ignorespaces #2}}   

\begin{document}
\bibliographystyle{apalike}

\title{\vspace{-15mm}\fontsize{24pt}{10pt}\selectfont\textbf{A comparison of priors for variance parameters in Bayesian basket trials}}

\author{
    Massimo Ventrucci \\
	Department of Statistical Sciences\\    
    University of Bologna\\
    \and
    Alessandro Vagheggini\\
   Clinical Safety Statistics, Biostatistics and Research Decision Sciences\\ 
   MSD Innovation and Development GmbH, Switzerland 
}
    
\maketitle

\begin{abstract}
Phase II basket trials are popular tools to evaluate efficacy of a new treatment targeting genetic alteration common to a set of different cancer histologies. 
Efficient designs are obtained by pooling data from the different arms (e.g., cancer histologies) via Bayesian hierarchical modelling, with a variance parameter controlling the strength of shrinkage of each arm treatment effect to the overall treatment effect. One critical aspect of this approach is that prior choice on the variance plays a major role in determining the strength of shrinkage and impacts the operating characteristics of the design. We review the priors most commonly adopted in previous works and compare them with the recently introduced penalized complexity (PC) priors. Our simulation study shows comparable behaviour for the PC prior and the gold standard choice {half-\emph{t}} prior, with the former performing better in the homogeneous scenario where all histologies respond similarly to the treatment. We argue that PC priors offer advantages over other priors because they allow the user to handle the degree of shrinkage by means of only one parameter and can be elicited based on clinical opinion when available.
\end{abstract}

\miscinfo{Keywords}{BHM; hierarchical logistic regression; INLA; PC prior; phase II trial}\\
\miscinfo{Address for Correspondence}{Massimo Ventrucci, Department of Statistical Sciences, University of Bologna. Email: massimo.ventrucci@unibo.it}


\section{Introduction}
\label{sec1}

Nowadays, cancer molecular characterization is rapidly changing oncology's therapeutic paradigm towards biomarker-driven treatments able to target specific genomic alterations. 
The previously conventional one-indication-fits-all is being replaced by \emph{precision medicine} which aims to target the right treatments to the right patients at the right time.
Clearly, drugs development must account for this therapeutic shift and their approval process has to cope with the new challenges. Classifications based on genomic alterations induce low prevalence, considerably reducing trial sample sizes. Basket trials allow to address this limitation by studying a new biomarker-driven treatment across multiple histologies. They can be seen as a collection of single-arm exploratory phase II studies, where the aim is to detect tumor histologies which may benefit from a new treatment targeting a communal genomic mutation \citep{renfro2017statistical,woodcock2017master}.

A simple and early strategy for detecting a treatment efficacy among groups of several cancer histologies sharing a common genomic alteration has been to statistically plan the study as a set of parallel separate non-randomized single-arm designs; e.g., adopting several Simon designs \citep{simon1989optimal}, one for each tumor histology. 
However straightforward, collections of separate studies do not account for any possible similarities in the response between tumors. This approach is known to alleviate possible bias, but at the same time may lead to loss of power especially in low sample size cases, which are frequent for cancer molecular characterization \citep{kane2019analyzing}

In this view, \citet{thall2003hierarchical} first proposed a Bayesian hierarchical modelling (BHM) approach for a phase II sarcoma trial with multiple subtypes, each corresponding to one arm, which allows borrowing strength of information (i.e., pooling data) between the different arms. This approach introduces a set of random effects to capture arm-specific drug responses and model them as independent random variables following a Gaussian distribution with mean $\mu$ and standard deviation $\sigma$. In this way, estimation of the efficacy rate (i.e., response rate) in a given arm will benefit from information coming from the other arms, being a compromise between the local (arm-specific) response rate  and the global (overall arms) response rate. The arm-level standard deviation $\sigma$ controls the strength of shrinkage of each arm response rate towards the overall efficacy rate.  Basket trials usually involve a small number of arms (e.g., four or five) hence the data carries little information about $\sigma$. As a consequence, the prior on $\sigma$ (or, equivalently, on the variance $\sigma^2$, or precision $\tau=1/\sigma^2$) will inevitably play a big role in determining the operational characteristics of the design. This makes the choice of prior in basket trials a major challenge for clinicians/practitioners.

\subsection{Previous work on the issue of prior choice in basket trials} 

We review previous work on the issue of prior choice on a classic BHM framework where the histologies are assumed as exchangeable. Hereafter, we will use the standard deviation $\sigma$ to generally refer to the arm-level variability. (However, we will present each prior in their most convenient scale, as it was firstly presented in the proponent papers; e.g. the conjugate Gamma on $\tau=1/\sigma^2$.)

In the seminal paper by \cite{thall2003hierarchical} a conjugate $Gamma(2, 20)$ on $\tau$ was used with the purpose of inducing moderate borrowing of information across arms. \cite{berry2013bayesian} proposed a weakly informative conjugate $Gamma(0.0005, 0.000005)$ in an attempt of preventing over-shrinkage and showed via simulation that this choice of prior overpowers a collection of Simon designs. In order to investigate the performance of BHM under different Gamma specifications, \cite{freidlin2013borrowing} conducted a thorough simulation study which led them to criticize the BHM approach to oncology phase II basket trials when the sample size is small or the number of tumor histologies is equal or less than five.

\cite{cunanan2019variance} went one step further by comparing the Gamma prior on $\tau$ with the Uniform and {half-\emph{t}} priors on $\sigma$. The last two have become popular after the work by  \cite{gelman2006prior} who advocated their use in hierarchical/multilevel Bayesian models where the number of groups is small. The empirical study by  \cite{cunanan2019variance} gives very useful insights and supports the points made by \cite{gelman2006prior}. Firstly, it was found that the operating characteristics obtained under the Gamma prior were highly variable across a range of scenarios, due to the high sensitivity to its parameters, shape and scale. Secondly, both the Uniform (with lower bound at $0$ and upper bound larger than $1$) and the {half-\emph{t}} (with scale parameter larger than $1$) shown desirable behaviour in terms of more stable operating characteristics. Their conclusion is that priors assigning large mass near $\sigma=0$ should be avoided as they force excessive shrinkage, pointing out that most Gamma specification have this property. Thus, they recommend the use of the Uniform or {half-\emph{t}} which guarantee substantial mass is placed in the tail (i.e., far from $\sigma=0$) grounded on the more robust operating characteristics under these priors. 

Finally, works related to the issue of prior choice has been done beyond the classic BHM framework. While the focus of our paper is on the classic BHM, we briefly report on previous work relaxing the assumption of exchangeability of the histologies in favour of more flexible models \citep{neuenschwander2016robust,hobbs2018bayesian}. 
Adopting an empirical Bayes approach,  \cite{chu2018bayesian} proposed to determine the amount of shrinkage across histologies as a transformation of a heterogeneity measure of the responses. Other strategies have focused on procedures which allow shrinkage only among homogeneous arms  \citep{leon2012borrowing,chu2018blast,chen2019bayesian,fujikawa2019bayesian,zheng2019borrowing} or subsequently to an interim analysis showing evidence in favor of homogeneity \citep{liu2017increasing}. Lastly,  \cite{psioda2019bayesian} suggested a Bayesian model averaging over the space of all the possible combinations of effective and ineffective histologies.

\subsection{Aim of the paper}

From \cite{cunanan2019variance} we learn two important facts. First, given its overall good performance the {half-\emph{t}} on the arm level standard deviation can be assumed as a sort of \emph{gold standard} for basket trial designs conducted via BHM. Second, the operating characteristics will essentially depend on the amount of mass concentrated near $\sigma=0$ assumed by the prior. Because the data itself is often scarcely informative in basket trials, this particular feature of the prior is what determines the strength of shrinkage to the overall treatment effect. 

We believe that a desirable prior in the context of a basket trial is one enabling the clinician/practitioner to set the amount of probability mass near $\sigma=0$ in an intuitive way. This elicitation process should be done according to the clinician belief about the level of shrinkage required in the trial. Building upon this motivation, this paper contributes to the literature by investigating the operating characteristics obtained under the Penalized Complexity (PC) priors recently proposed by \cite{simpson2017penalising}. By definition a PC prior is an exponential with rate parameter $\lambda$, defined on a scale measuring the distance from a well defined \emph{base model}. In a classic BHM context, a natural base model is the one where the response rate is constant across arms (i.e., $\sigma=0$, which we denote as the homogeneous scenario). If we assume a PC prior with base model $\sigma=0$, then $\lambda$ acts as an (hyper-)parameter that controls directly the strength of shrinkage to the overall treatment effect. Importantly, the degree of shrinkage can be tuned in a monotonic way: as $\lambda$ increases, more and more mass is placed near the base model $\sigma=0$, thus enforcing shrinkage. For instance, a large $\lambda$ may be used when the clinician anticipates homogeneity in the response rate across the arms, while moderate or small $\lambda$ may be chosen in heterogeneous cases where strong shrinkage is not required. Thus, PC priors seem to offer a potential advantage over the Uniform and {half-\emph{t}}, in that the user will have to handle only one parameter to tune the strength of shrinkage and control the properties of the design. 

In order to realize the practical advantages offered by PC priors, some intuition about the scale of $\lambda$ must be provided. The first goal of this paper is to propose intuitive methods to elicit PC priors (i.e., choose $\lambda$) based on clinical opinion. Our aim is to find methods that can help practitioners to translate clinical information (e.g., the anticipated degree of homogeneity of the histologies) into a value for $\lambda$. 
Our second goal is to get insights about the operating characteristics attained by PC priors as compared to the gold standard {half-\emph{t}}. This will permit to identify scenarios where PC priors lead to more efficient designs. This goal is addressed by a simulation study which evaluates the operating characteristics obtained under several popular prior choices in a range of scenarios, which we believe cover typical basket trial settings. The main focus of the simulation will be on the comparison between the PC prior and the {half-\emph{t}}. 

An accompanying R package called \texttt{INLAbhmbasket} is produced as a tool to support practitioners in planning phase of basket trials. The package provides tools for simulation of basket trials and computation of the operating characteristics under several choices of priors (Gamma, Uniform, PC prior and {half-\emph{t}}) and specification of the study (number of patients, cutoff probabilities, accrual rates, etc). 
(The \texttt{INLAbhmbasket} package is available at \url{https://github.com/massimoventrucci/INLAbhmbasket}. R code to simulate basket trials under various prior choices and compute operational characteristics can be found in Appendix \ref{app:Rcode}).

The plan of the paper is as follows. In Section~\ref{sec:basket} the sequential Bayesian design strategy for phase II oncology basket trials and the BHM approach are described in detail. In Section~\ref{sec:priors} the most popular prior choices for variance parameters in hierarchical models are reviewed, with particular emphasis on the PC prior framework. Section~\ref{sec:scaling_pc} describes methods to choose the PC prior parameter $\lambda$ in the context of basket trials. 
In Section~\ref{sec:sim}, results from our simulation study are presented. The paper ends with a discussion in Section~\ref{sec:discussion}.

\section{Bayesian basket trials}
\label{sec:basket}

Phase II basket trials aim to estimate the efficacy rate of a new treatment targeting a specific genomic alteration common to a set of $J$ different tumor histologies.
Let $y_j$ be the observed count of patients who positively respond to the treatment, the observation model is
\begin{eqnarray}
\label{eq:data_model} 
y_j & \sim & \text{Binomial}(N_j, p_j) , \quad j = 1, \ldots, J,
\end{eqnarray}
where $p_j$ and $N_j$ are respectively the true response rate (efficacy rate) and sample size for histology $j$. For each histology, we are interested in testing the null hypothesis $H_0: p_j \leq q_{0}$ versus the alternative one $H_1: p_j \geq q_{1}$, where $q_{0}$ and $q_{1}$ represent respectively uninteresting and desirable target levels for the response rate (usually $q_{1} - q_{0} = 0.15, 0.20, 0.25$). 

\subsection{Sequential Bayesian design}
\label{sec:design}


To increase the ethical component of the trial a sequential design is adopted. At each step, any arm can be closed due to futility or efficacy. To account for possible varying patients accrual rates across arms (e.g., one cohort might enroll patients quicker than the others), the first interim analysis is performed after a fraction $0 < \omega < 1$ of each arm maximum sample size $\lceil \omega N_j \rceil$ is enrolled (where the ceiling operator $\lceil x \rceil$ represents the least integer greater than or equal to $x$). Let $n_j^{(1)} \geq \lceil \omega N_j \rceil$ ($j = 1, \ldots, J$) represent the actual number of patients enrolled in arm $j$ at the first halt; then, $n^{(1)} = \sum_{j = 1}^J n_j^{(1)}$ is the corresponding total number of patients and $y^{(1)} = (y_1^{(1)}, \ldots, y_J^{(1)})^\top$ the related vector of data observed up to the first stop.
Following \cite{berry2013bayesian} at each interim analysis both futility and efficacy are assessed in each arm. Accrual in arm $j$ is stopped for futility if
\begin{equation}
\label{eq:futility_stop}
\Pr (p_j > \bar{q}\, | \, y^{(1)}) < 0.05
\end{equation}
or efficacy if
\begin{equation}
\label{eq:efficacy_stop}
\Pr (p_j > \bar{q}\, | \, y^{(1)}) > 0.90,
\end{equation}
where these probabilities are computed on the basis of the estimated posterior distribution of the response rate $p_j$. Midpoint $\bar{q} = (q_{0} + q_{1}) / 2$ is chosen instead of $q_{0}$ as a proof beyond reasonable doubt to deem the treatment futile/active at early stages. The next possible halts take place at steps of $\lceil k n^{(1)} \rceil$ patients, where $0 < k < 1$ is conveniently chosen to manage the frequency of the interim analyses.
At each of these stops, rules (\ref{eq:futility_stop}) and (\ref{eq:efficacy_stop}) are applied on the accumulated data $y^{(l)}$ ($l > 1$).
Enrollment resumes only in those arms deemed neither futile nor active. The trial ends when all the arms have been closed or the preset maximum sample size has been reached.
The final analysis is based on the whole accrued data $y$ and the treatment will be declared effective for the histology $j$ if
\begin{equation}
\label{eq:final_analysis}
\Pr(p_j > q_{0} \, | \, y) > \zeta, \quad \quad j = 1, \ldots, J,
\end{equation}
where $\zeta_j$ is a probability cutoff ensuring type I error control at a preset level $\alpha$ \citep{berry2011bayesian}.

It is worth stressing that Bayesian sequential monitoring, as is the case for rules (\ref{eq:futility_stop}) and (\ref{eq:efficacy_stop}), is not affected by any multiplicity issue contrary to the frequentist approach \citep{berry1993case}.

\subsection{Bayesian hierarchical modelling (BHM)}
\label{sec:bhm}

The BHM approach relies on a hierarchical model where at the first level we have the observation model in (\ref{eq:data_model}).
At the second level, the linear predictor $\eta_j = \operatorname{logit}(p_j) = \log(p_j / (1 - p_j))$ is modelled as
\begin{eqnarray}
\eta_j| \mu, \sigma^2 & \sim & \text{Normal}(\mu, \sigma^2) \quad \quad j = 1, \ldots, J.
\label{eq:lin_pred}
\end{eqnarray}
The modelled quantity $\eta_j$ is the log-odds of the response rate. In some works \citep{thall2003hierarchical,berry2013bayesian,cunanan2019variance} an offset is included in the linear predictor and the modelled quantity is the logit deviation from the target $q_1$, $\operatorname{logit}(p_j) - \operatorname{logit}(q_{1})$. The inclusion of an offset has no implications on the posterior distribution of the probability values $p_j$'s, which are the quantities of interest in the sequential design described in Section~\ref{sec:design}. 

The model in (\ref{eq:lin_pred}) depends on two hyperparameters, $\mu$ and $\sigma^2$. The former is the overall response to the treatment expressed in the logit scale. An uninformative prior $\mu \sim \text{Normal}(0, 100)$ is a common choice to express uncertainty about the overall efficacy rate. The variance $\sigma^2$ is the hyper-parameter controlling pooling of information across arms. A small value of $\sigma$ favours pooling, while a large $\sigma$ favours locality and returns arm-specific estimates less shrunk towards $\mu$. For this reason, the choice of priors on $\sigma$ is a critical choice to be made by the user. 
Popular strategies to choose this prior will be discussed in Section~\ref{sec:priors}. 

For practical purpose we reparameterize (\ref{eq:lin_pred}) as $\eta_j = \mu + \theta_j$ with priors
\begin{equation}
\mu  \sim  \text{Normal}(0,100) 
 \quad ; \quad \theta_j | \sigma^2  \sim  \text{Normal}(0, \sigma^2) \quad \quad j = 1, \ldots, J.
\label{eq:lin_pred_rep}
\end{equation}
We find this reparameterization more convenient in practice, because model (\ref{eq:lin_pred_rep}) can be implemented straightforwardly in the R package R-INLA \citep{rue2009approximate} for approximate Bayesian inference. The INLA approach is computationally more efficient than MCMC, a feature that turns out to be very helpful in designing basket trials via BHM. Checking the impact of design's parameters, like the (hyper-)parameters in the prior specification for $\sigma$, requires estimating the model hundreds of times to evaluate the operating characteristics under several scenarios, and this can be done relatively quickly by using INLA.

\section{Prior choice on the arm-level variance}
\label{sec:priors}

The task of prior choice on $\sigma$ would be greatly simplified if some information about the degree of homogeneity of the histologies were available at prior. 
We can distinguish between two opposite scenarios: \emph{homogeneous trials} (i.e., $\sigma=0$), where either all arms are active (positive response to the treatment) or all are inactive (treatment is ineffective), \emph{heterogeneous trials} (i.e., $\sigma>0$), where some of the arms are active while others are not.  The operating characteristics achieved by a certain prior on $\sigma$ largely depend on whether the trial is homogeneous or heterogeneous. A prior assigning high probability mass near $\sigma=0$ will represent a suitable choice in homogeneous scenarios, as this prior favours pooling. A prior distributing more mass away from $\sigma=0$ (i.e., in the tail of the distribution) will be appropriate in heterogeneous scenarios as this prior favours locality. 

It is useful to discuss the consequences, in terms of power detection and type I error control, of choosing a prior that favours pooling (prior A) as opposed to one that favours locality (prior B).  In homogeneous trials with all active arms, prior A will guarantee high power detection. However, in heterogeneous trials, prior A will incur in \emph{over-shrinkage} causing loss of power detection on the active arms and inflated type I error rate on the inactive ones. Regarding prior B, in heterogeneous trials this will probably guarantee both reasonable power on the active arms and type I error control on the inactive ones. However, in homogeneous trials with all active arms, prior B will incur in \emph{over-fitting} leading to reduced power detection, hence an inefficient design. 

As we can see, the inevitable trade-off between high power detection and strict type I error control is linked to two factors: first, the \emph{balance between poling and locality} implied by the prior on $\sigma$ and, second, the specific features of the trial under study. One important feature is the relative importance of high power and strict type I error control, which can vary across studies according to their primary goal. The other relevant aspect regards the level of homogeneity/heterogeneity of the trial. In principle, the practitioner must choose the appropriate balance between pooling and locality, according to the information available on the given trial.

In general, information about homogeneity of the trials will unlikely be available at prior, given the early development stage of a phase II basket trial. Nevertheless, in some cases clinicians may leverage their experience from past studies in order to make a prior guess on the homogeneity of the trial at hand.  Thus, we believe that a prior on $\sigma$ that enables intuitive control of the balance between pooling and locality would be an important tool for clinicians involved in designing basket trials; including the possibility to be used as default choice in some studies. Below, we discuss some popular classes of priors commenting on how easily the involved parameters can handle pooling versus locality.

\subsection{Conjugate Gamma}

A popular prior for the scale parameter $\sigma^2$ is the conjugate Inverse-Gamma$(a, b)$ which corresponds to a Gamma$(a, b)$ on the precision $\tau=1/\sigma^2$, \[\pi(\tau|a,b) \propto \tau^{a-1} \exp(-b \tau),\] 
where $a$ and $b$ are respectively the shape and rate parameters. 
The conjugate Gamma has been criticized in several papers as a prior forcing over-fitting \citep{FruhwirthSchnatter-2010,FruhwirthSchnatter-2011,simpson2017penalising}. For our purposes, we note that it is not immediate to outline simple strategies to define increasing/decreasing levels of shrinkage by handling parameters $a$ and $b$. Thus, while control of the balance between pooling and locality is possible, the Gamma fails to provide the practitioner with practical and intuitive ways to do so. 
In our simulation study we will only focus on the specification with $a=0.0005$ and $b=0.000005$ proposed in a previous work by \cite{berry2013bayesian}.

\subsection{Half-t}
The {half-\emph{t}} on the standard deviation $\sigma$ has been popularized in Bayesian hierarchical models by \cite{gelman2006prior}. We denote this distribution as {half-\emph{t}}$(\gamma, \nu)$, where $\gamma$ is the scale parameter and $\nu$ is the number of degrees of freedom. The density is given by
\[
\pi(\sigma|\gamma,\nu) \propto \left( 1+\frac{1}{\nu}\left(\frac{\sigma}{\gamma}\right)^2 \right)^{-\frac{\nu+1}{2}}.
\] 
For $\nu=1$ we obtain the half-Cauchy prior, while for $\nu=-1$ we have the improper Uniform. In a thorough simulation study, \cite{cunanan2019variance} demonstrated robustness of the {half-\emph{t}} in the context of BHM of basket trials. In our simulation study we will only focus on the specification with $\nu=1$ and $\gamma=10$, based on results from \cite{cunanan2019variance}.

The user can control balance between pooling and locality by manipulating $\nu$ and $\gamma$. However, there is not a unique strategy to do. For instance, to place more and more mass in the tail of the distribution, which would give a prior that favours locality, one could either increase $\gamma$ while fixing $\nu$, or decrease $\nu$ while fixing $\gamma$. In our view, the need to handle two parameters, $\nu$ and $\gamma$, and the lack of a unique approach to tune pooling versus locality represent impractical features of the {half-\emph{t}} in the context of basket trials.

\subsection{Uniform}

The uniform distribution $U(a,b)$ assigns constant probability in the interval $(a,b)$, with density $\pi(\sigma) = 1/(b-a)$ for $a \leq \sigma \leq b$, while $\pi(\sigma)=0$ elsewhere. By manipulating $a$ and $b$ the user can tune the probability mass concentrated near $\sigma=0$. For instance, by increasing the upper bound $b$ while fixing $a=0$, increasing mass is assigned to the tail of the distribution. However, a prior that favours locality can also be achieved by playing with the lower bound $a$ by setting it to a value larger than $0$ \citep{cunanan2019variance}. 

Analogously to the {half-\emph{t}} case, with the uniform the user has no unique strategy to control balance between pooling and locality, which, again, we see as an inconvenient feature in basket trials.

\subsection{PC prior}

The PC prior by \cite{simpson2017penalising} is built under a principled framework that provides the user an intuitive way to control/constrain model complexity. The prior is defined on a scale measuring the distance from the base model via the Kullback-Leibler divergence \citep{kld-1951}.
For the sake of comparison with the {half-\emph{t}} prior, we report the PC prior for the standard deviation $\sigma$, which is
\begin{eqnarray}
    \pi(\sigma) 
&=& \lambda \exp(-\lambda \sigma),
    \label{eq:pcprior_sigma}
\end{eqnarray}
where $\lambda$ is the rate of the exponential distribution; for more details see Appendix \ref{app:pc}. 

For the PC prior (\ref{eq:pcprior_sigma}) we have a unique strategy to balance pooling versus locality through the rate parameter $\lambda$, which controls the penalty for deviating from the base model $\sigma=0$. The tuning mechanism is monotonic: large values of $\lambda$ concentrate more mass at the base model, while small values of $\lambda$ distribute more mass in the tail of the distribution. The advantage is that by means of one parameter, $\lambda$, the user can control directly the strength of shrinkage to the overall treatment effect, which yields a very practical approach.

\section{Scaling the PC prior: choice of $\lambda$} 
\label{sec:scaling_pc}
We describe two possible user-defined scalings (i.e., ways to select $\lambda$) of the PC prior which we believe may be of practical use for the clinician who has some prior information on the specific basket trial under examination. 

\subsection{Scaling 1: choosing $\lambda$ based on a guess on the standard deviation of $\theta_j$}
\label{sec:scaling1}

The choice of $\lambda$ can be done in practice by eliciting a statement of the form $\Pr(\sigma > z) > c$ \citep{simpson2017penalising}. Let $c$ be a small probability, then $z$ can be regarded as an upper bound for $\sigma$. It can be shown that given certain $c$ and $z$ chosen by the practitioner, then $\lambda={-\ln(c)}/{z}$.

\cite{simpson2017penalising} suggested a practical rule of thumb to choose $\lambda$ based on the above criteria, which we find useful in the context of basket trials. This requires the user to elicit a guess on the marginal standard deviation of $\theta_j$ (i.e., the arm $j$ deviation from the overall treatment effect $\mu$) expressed on the logit scale from Eq. (\ref{eq:lin_pred_rep}). This scaling approach is based on the following argument: let assume a PC prior on $\sigma$ with rate parameter $\lambda={-\ln(c)}/{z}$, then the marginal standard deviation of $\pi(\theta)= \int \pi(\theta_j|\sigma)\pi(\sigma) d\sigma$ - after marginalizing out the uncertainty on $\sigma$ - is approximately $0.31z$, when $c=0.01$ \citep{simpson2017penalising}. The implied rule of thumb is: let $sd$ be the practitioner's guess on the marginal standard deviation of $\theta_j$, then
\begin{equation}
\lambda={-0.31\ln(0.01)}/{sd},
\label{eq:lambda_sd}
\end{equation}
hence the choice of $\lambda$ amounts to select a suitable value of $sd$. 

\begin{figure}[htb]
\centerline{
\includegraphics[scale=0.6]{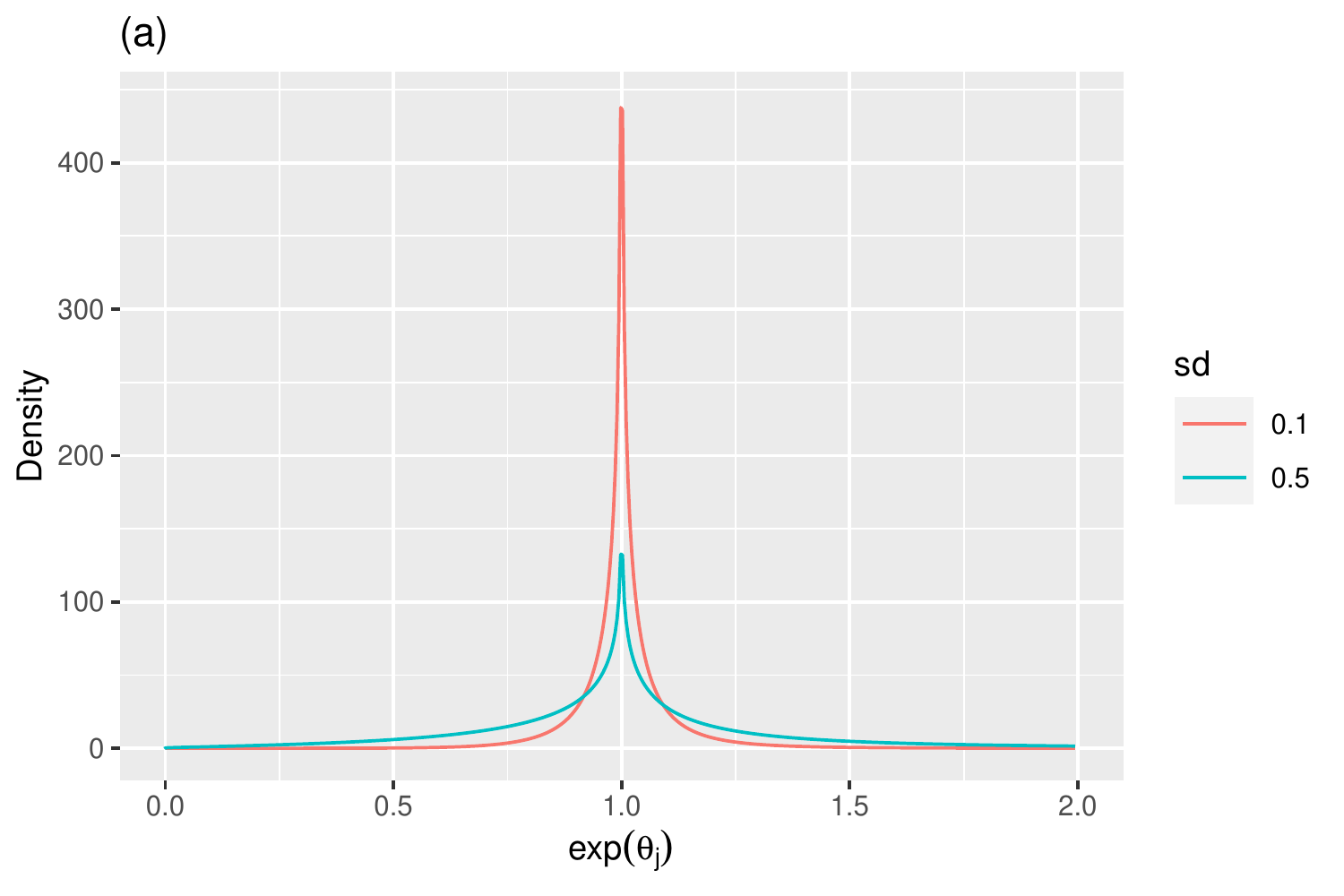}
\includegraphics[scale=0.6]{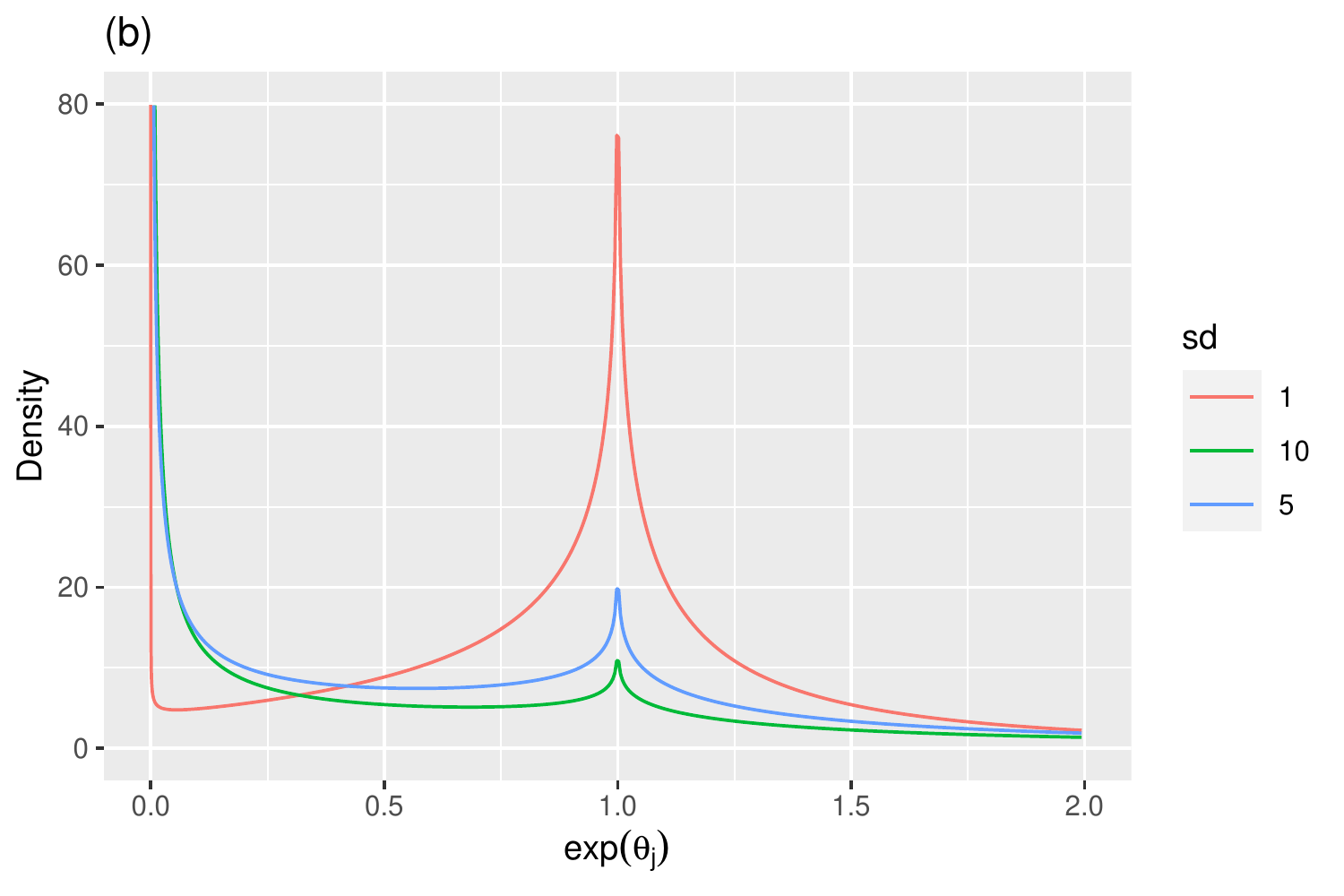}}
\caption{The prior on $\exp(\theta_j)$ (odds ratio) implied by using a PC prior on $\sigma$, for various choices of $sd$. In panel (a), we explore the case where $sd$ is set to small values (e.g., $sd =\{0.1,0.5\}$). In panel (b), we look at the prior implied by a larger value $sd = \{1,5,10\}$. In both panels the prior shows a peak at $1$, which is the odds ratio when the arm-$j$ treatment effect equals the overall  treatment effect.
}
\label{fig:impliedor}
\end{figure}

The practitioner may leverage clinical opinion or external/past data to define a suitable value of $sd$ for the trial at hand. One way to address the choice of $sd$ in practice is to ask, for instance, what $sd=1$ means in terms of an easy-to-interpret transformation of $\theta_j$, such as the odds ratio $\exp(\theta_j)$. Figure \ref{fig:impliedor} shows the \emph{implied prior} on the odds ratio $\exp(\theta_j)$, corresponding to using a PC prior on $\sigma$ with various $sd$ values. (This prior is obtained by numerically computing $\pi(\theta_j) = \int \pi(\theta_j|\sigma)\pi(\sigma)d \sigma$, then applying a change of variable to derive the prior on the odds ratio scale $\exp(\theta_j)$.) Note that all the priors in Figure \ref{fig:impliedor} peak at $1$ which is the value of the odds ratio when the arm-$j$ treatment effect equals the overall treatment effect. Essentially, at base model $\sigma=0$ we have $\exp(\theta_j)=1, j = 1, \ldots, J$. Also note that the priors show different decay from the base model according to $sd$. From Figure \ref{fig:impliedor}(a), when a small $sd$ is chosen (e.g., $<1$) the prior is narrowly concentrated around $1$, hence variation in the odds ratio is very small. For instance, the PC prior with $sd=0.1$ (i.e., $\lambda= -0.31\ln(0.01)/0.1$) places most of the mass inside the interval $(0.8,1.2)$; thus, the choice $sd=0.1$ would be coherent with a user anticipating around $20\%$ increase/decrease in the arm-specific odds ratio. This prior choice is hardly desirable in practice as it will induce very strong shrinkage. A more moderate degree of shrinkage is induced by priors in figure \ref{fig:impliedor}(b), where we can appreciate that larger values of $sd$ still lead to substantial shrinkage but, at the same time, allow more variability in the odds ratios. As an example, a user setting $sd=1$ is implying that odds ratios can vary approximately in the range $(0.5,1.5)$. This choice may be suitable in trials where a moderate/large degree of homogeneity between the histologies is anticipated. Finally, $sd=5$ and $sd=10$ will induce low level of shrinkage, resulting in much larger variability in the odds ratio. This choice may be suitable when an heterogeneous trial is anticipated.

In our simulation study we will test the performance of the PC prior for $sd=(1,5,10)$.


\subsection{Scaling 2: choosing $\lambda$ by matching the tail of the {half-\emph{t}}}

 
We propose a second scaling approach which aims at matching the tail properties of a given {half-\emph{t}}$(\gamma,\nu)$ distribution. In this case, the user is only required to select the parameters $\gamma$ and $\nu$ of the {half-\emph{t}} prior that they want to reproduce, then Proposition \ref{PROP:LAMBDA} below tells what $\lambda$ has to be for a PC prior having approximately the same tail behaviour. Precisely, same tail behaviour means same probability mass assigned to a \emph{tail interval} of the form $[x, \infty)$, $x > 0$. We will denote such PC prior as \emph{equivalent PC prior} (EPC), meaning that it is equivalent to the {half-\emph{t}} in terms of the probability mass assigned to the tail interval.

\begin{proposition}\label{PROP:LAMBDA}
The PC prior in Eq. (\ref{eq:pcprior_sigma}) with rate parameter given by \begin{equation}
\label{eq:lambda}
\lambda_{\gamma, \nu}(x) = - \ln \left[ 1 - I \left( \frac{x^2}{x^2 + \gamma^2 \nu}; \frac{1}{2}, \frac{\nu}{2} \right) \right] / x.
\end{equation}
will have the same probability mass as the {half-\emph{t}}$(\gamma,\nu)$ in the interval $A = [x, \infty)$, $x > 0$. (Notation $I ( \cdot; \cdot, \cdot)$ indicates the beta regularized function \citep{abramowitz1965handbook}). 
\end{proposition}

\begin{proof}
See Appendix \ref{app:proof}.
\end{proof}

Figure~\ref{fig:lambda} displays $\lambda_{\gamma, \nu}(x)$ in (\ref{eq:lambda}) as a function of  $x$ (i.e., the lower bound of the tail interval $A$). It can be shown that as $x \rightarrow \infty$, then $\lambda_{\gamma, \nu}(x)$ goes to $0$ and the resulting PC prior becomes increasingly flat. As $x \rightarrow 0^+$,  then $\lambda_{\gamma, \nu}(x)$ goes to $\lambda_0 = 2( \gamma \nu^{1 / 2} \, B ( 1 / 2, \nu / 2 ) )^{- 1}$, where $B(\cdot, \cdot)$ is the beta function. What is important to notice is the non-monotonic behaviour of $\lambda_{\gamma, \nu}(x)$, which results problematic because two different tail intervals $A=[x, \infty)$ may be associated to the same value of $\lambda_{\gamma, \nu}(x)$. The definition of tail interval must be made unambiguous in order to uniquely define the EPC. We then introduce the idea of a reference tail interval $A^*$, which is defined as the largest $A=[x, \infty)$ such that $\lambda_{\gamma, \nu}(x) \leq \lambda_0$; the value of $x$ satisfying the condition above can be found numerically. Note, for any choice of the parameters of the {half-\emph{t}} we can get the associated reference tail interval $A^*$. Based on $A^*$, the concept of equivalent PC prior can be defined.

\begin{remark}
\label{def:epc}
We define as \emph{equivalent PC prior} with scale parameter $\gamma$ and degrees of freedom $\nu$, denoted as EPC$_{\gamma,\nu}$, the PC prior in Eq. (\ref{eq:pcprior_sigma}) with rate parameter $\lambda=\lambda_{\gamma, \nu}(x)$, where $\lambda_{\gamma, \nu}(x)$ is obtained from Proposition \ref{PROP:LAMBDA} using $A=A^*$.
\end{remark}

Table~\ref{tab:lambda_htd} reports $\lambda_{\gamma, \nu}$ for different specifications of the {half-\emph{t}}. As an example, the PC prior with $\lambda=0.064$ is an EPC$_{10,1}$. This means that the EPC$_{10,1}$ and the {half-\emph{t}}$(\gamma=10,\nu=1)$ assign the same probability mass to the reference tail interval $A^*$.


\begin{table}[htb]
\centering
\caption{The value of $\lambda_{\gamma,\nu}$ for various parameters of the {half-\emph{t}} with scale parameter $\gamma$ and degrees of freedom (dof) $\nu$. Between brackets the corresponding $sd$ computed from (\ref{eq:lambda_sd}).} 
\label{tab:lambda_htd}
\tabcolsep=0.1cm
\begin{tabular}{c c c c c c c}
\hline
& & \multicolumn{5}{c}{scale parameter $\gamma$} \\
\cmidrule{3-7}
  dof $\nu$  & & 1 & 2 & 5 & 10 & 20\\
\cmidrule{1-1}
\,~1 & & 0.637 \textit{(2.242)} & 0.318 \textit{(4.485)} & 0.127 \textit{(11.212)} & 0.064 \textit{(22.425)} & 0.032 \textit{(44.849)} \\
\,~2 & & 0.707 \textit{(2.019)} & 0.354 \textit{(4.038)} & 0.141 \textit{(10.095)} & 0.071 \textit{(20.189)} & 0.035 \textit{40.379)} \\
\,~5 & & 0.759 \textit{(1.880)} & 0.380 \textit{(3.761)} & 0.152 \,~\textit{(9.402)} & 0.076 \textit{(18.804)} & 0.038 \textit{(37.607)} \\
\,10 & & 0.778 \textit{(1.834)} & 0.389 \textit{(3.669)} & 0.156 \,~\textit{(9.172)} & 0.078 \textit{(18.345)} & 0.039 \textit{(36.689)} \\
\hline
\end{tabular}
\end{table}


\begin{figure}[htb]
\begin{center}
\includegraphics[scale = 0.75]{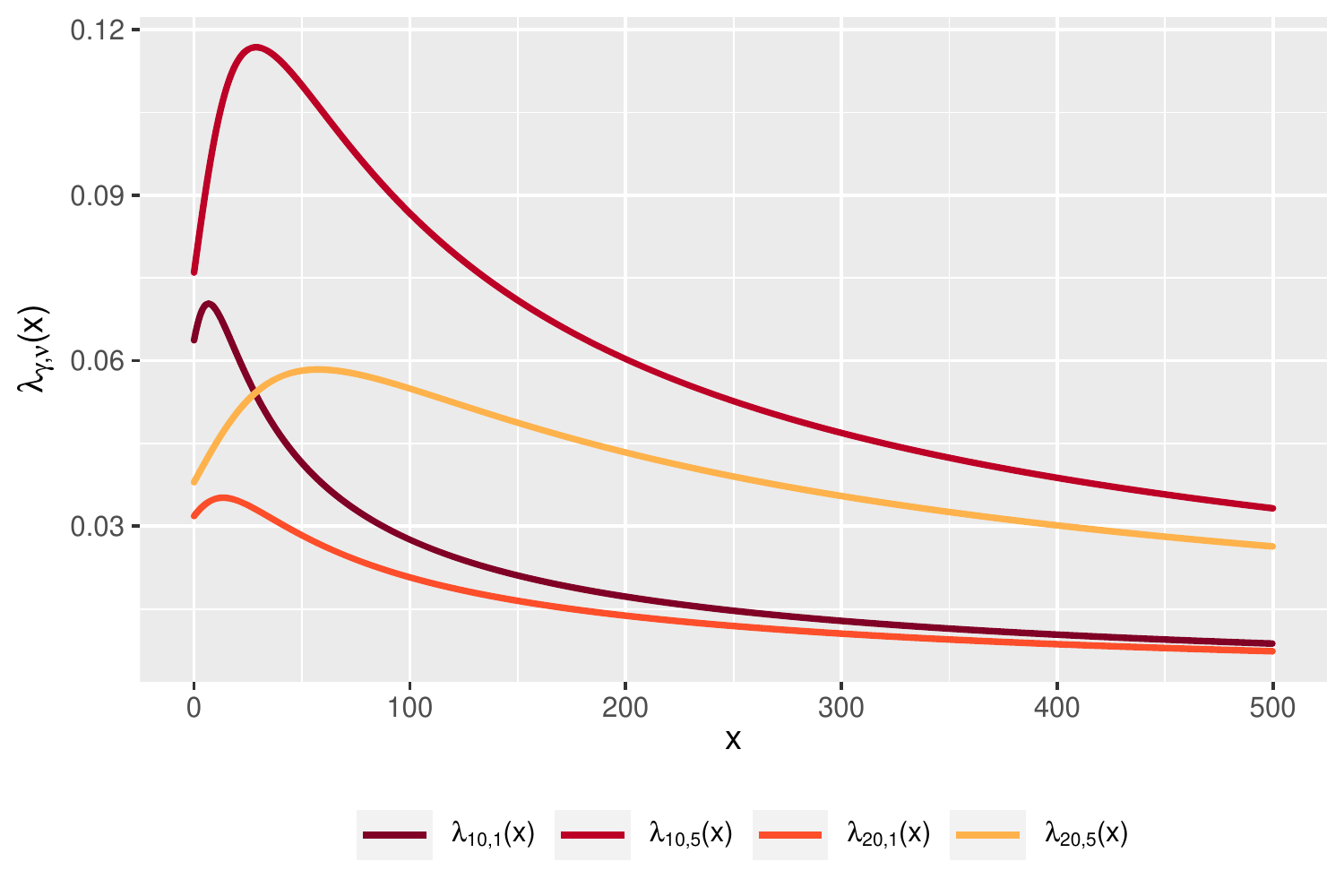}
\caption{Behaviour of $\lambda_{\gamma, \nu}(x)$ for different values of the {half-\emph{t}} scale parameter $\gamma$ and degrees of freedom $\nu$.}
\label{fig:lambda}
\end{center}
\end{figure}

\subsection{Comparing the two scaling approaches}

\begin{figure}[htb]
\begin{center}
\includegraphics[scale = 0.75]{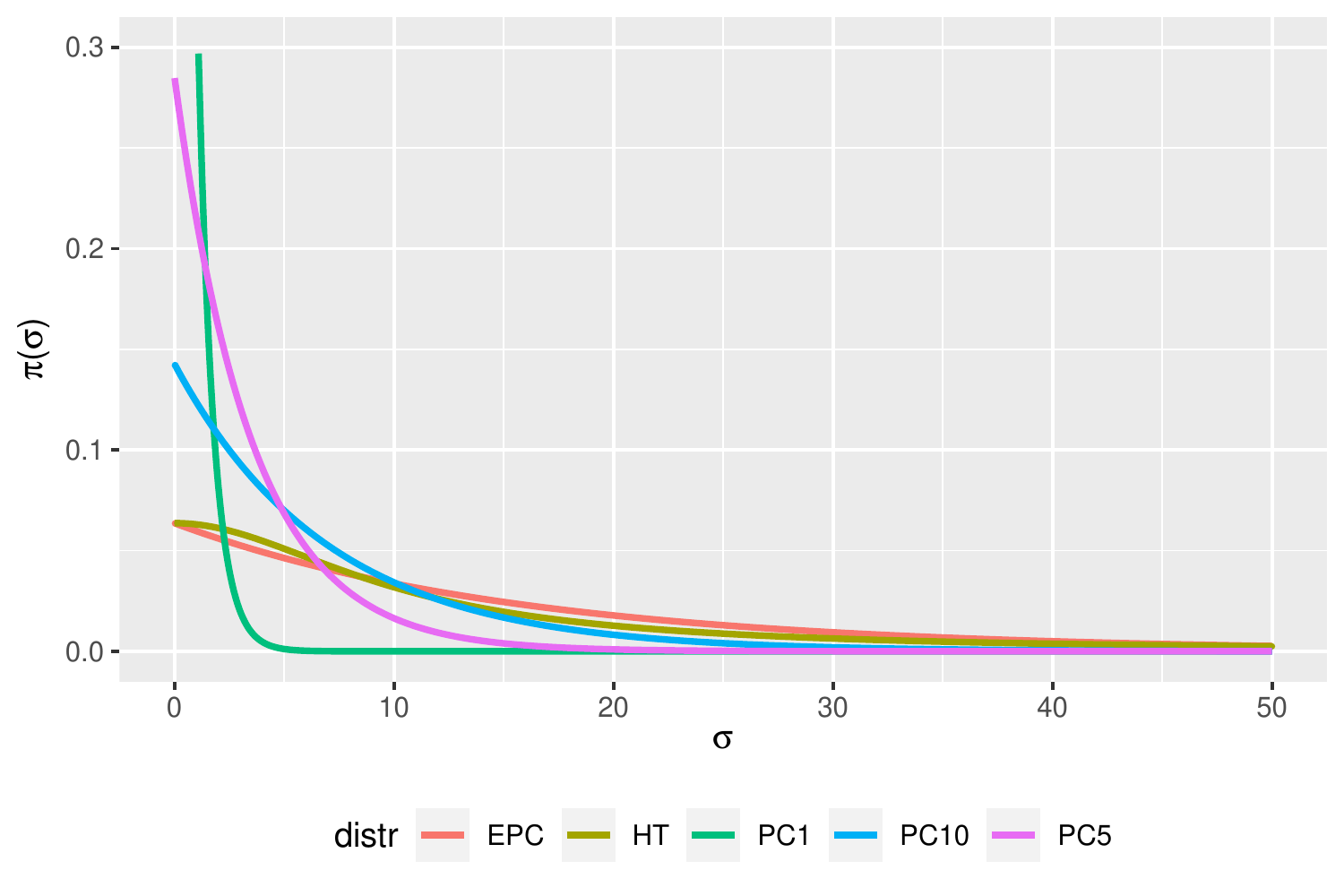}
\caption{Half-\emph{t} (blue) and PC prior (several shades of green) distributions on the standard deviation $\sigma$.}
\label{fig:priors}
\end{center}
\end{figure}

Let us compare the proposed two scaling approaches . Figure~\ref{fig:priors} displays the {half-\emph{t}} (HT), the PC prior with $sd =1, 5, 10$, denoted as PC1, PC5 and PC10, the {half-\emph{t}}$(\gamma = 10,\nu = 1)$ and the associated EPC$_{10,1}$. All distributions are expressed in the scale of the standard deviation to emphasize contraction towards the base model $\sigma=0$ (i.e., the level of shrinkage). By increasing $\lambda$ (i.e., decreasing $sd$) the level of shrinkage increases. The EPC$_{10,1}$ and the {half-\emph{t}}$(\gamma=10,\nu=1)$ have roughly the same tail behaviour, they only differ near $\sigma=0$: the PC prior goes to $\sigma=0$ exponentially, while the {half-\emph{t}} has bell-shaped behaviour. 

Figure~\ref{fig:priors} suggests that PC prior is a flexible class. For an appropriate choice of $\lambda$, the PC prior can reproduce the \emph{heavy tail} of the {half-\emph{t}} fairly well. In view of their adaptability to different scenarios, PC priors may be assumed as default priors in basket trials. The two different scaling approaches proposed may be used for different purposes. With scaling 2, the PC prior can potentially gain the stable operational characteristics achieved by the gold standard {half-\emph{t}} and this may be desirable in heterogeneous scenarios. With scaling 1, the $sd$ value can be tuned by the practitioner according to the strength of shrinkage required by the study. Prior information on the expected odds ratio can guide the choice of $sd$. In particular, scaling 2 is advantageous when clinicians anticipate homogeneity, in which case setting $sd=1$ can lead to potentially high power detection of the active arms.


\section{Simulation study}
\label{sec:sim}

We evaluate via simulation the frequentist operating characteristics attained by the sequential procedure described in Section~\ref{sec:design} under several priors in different scenarios. Our first goal is to investigate how the recently introduced PC priors compare to other priors previously analyzed, especially to the popular {half-\emph{t}} prior. The second goal is to provide guidelines for clinicians interested in using PC priors for BHM of basket trials on the most appropriate choice of the scaling parameter $\lambda$, in the different scenarios.

\subsection{Simulation scenarios}
Numerical results are based on 1000 four-armed simulated trials where the aim is to evaluate the null $H_0: p_j \leq 0.20$ against $H_1: p_j \geq 0.35$ at a $\alpha = 0.10$ significance level.
Data have been generated under five scenarios (reported in Table~\ref{tab:scenarios_interim}) each with an increasing number of active arms. In scenario 1, all arms are inactive and responses are generated under the null hypothesis (i.e., uninteresting response rate), while scenarios from 2 to 4 represent heterogeneous cases; in scenario 5, responses are all generated under the alternative hypothesis.


\begin{table}[htbp]
\centering
\caption{Simulation Scenarios; true values of $p_j$ in each arm are reported for each scenarios.} 
\label{tab:scenarios_interim}
\begin{tabular}{l c c c c}
\toprule
& \multicolumn{4}{c}{true $p_j$} \\
\cmidrule{2-5}
\multicolumn{1}{l}{scenario} & arm 1 & arm 2 & arm 3 & arm 4 \\ 
\cmidrule{1-1}
1. all null ($H_0$) & 0.20 & 0.20 & 0.20 & 0.20 \\
2. three null, one alternative & 0.20 & 0.20 & 0.20 & 0.35 \\
3. two null, two alternative & 0.35 & 0.20 & 0.20 & 0.35 \\
4. one null, three alternative & 0.35 & 0.35 & 0.20 & 0.35 \\
5. all alternative ($H_1$) & 0.35 & 0.35 & 0.35 & 0.35 \\
\bottomrule
\end{tabular}
\end{table}

Three maximum sample size levels common to all arms have been taken into account, $N_j = 20, 26, 37$. The sequential Bayesian design in Section~\ref{sec:design} has been employed with $\omega = 0.4$ and $k = 0.5$. Futility and efficacy are evaluated on the posterior distributions estimated at each interim analysis as defined in (\ref{eq:futility_stop}) and (\ref{eq:efficacy_stop}). Final analyses are based on (\ref{eq:final_analysis}) with probability cutoffs $\zeta$ \emph{ad hoc} determined, so to have the type-I error rate broadly equal to $\alpha=0.1$ in the null scenario 1. Table~\ref{tab:cutoffs} reports the $\zeta$ values, for each prior choice and sample size. For low sample sizes, cutoffs for the Gamma and Uniform were not identified because of numerical instability of the model implementing these priors; we did not experience any problem with the {half-\emph{t}} and the PC prior.


\begin{table}[htbp]
\centering
\caption{Probability cutoffs. In small sample size scenarios, we do not consider priors B and U due to computational instabilities of the model output under these prior choices.} 
\label{tab:cutoffs}
\begin{tabular}{c c c c}
\toprule
& \multicolumn{3}{c}{\bf $\zeta$} \\
\cmidrule{2-4}
prior & $N_j = 20$ & $N_j = 26$ & $N_j = 37$ \\
\cmidrule{1-1}
G & -- & -- & 0.878 \\
U & -- & -- & 0.866 \\
HT & 0.873 & 0.868 & 0.863 \\
PC1 & 0.859 & 0.859 & 0.852 \\
PC5 & 0.867 & 0.865 & 0.860 \\
PC10 & 0.871 & 0.867 & 0.862 \\
EPC & 0.871 & 0.867 & 0.863 \\
\bottomrule
\end{tabular}
\end{table}

As competitors of the PC prior we have selected the priors listed below which have been suggested in previous works:
\begin{itemize}
\item The Gamma$(0.0005, 0.000005)$ on $\tau=1/\sigma^2$ originally proposed by \cite{berry2013bayesian} (denoted as G);
\item The {half-\emph{t}} with scale $\gamma = 10$ and degrees of freedom $\nu = 1$ (denoted as HT) and the Uniform $U(0, 100)$ (denoted as U), which are two priors tested in \cite{cunanan2019variance} achieving robust results.
\end{itemize}

Regarding PC priors, both scaling approaches have been taken into account 
\begin{itemize}
\item Scaling 1: we consider PC priors with $sd=1,5,10$ (denoted as PC1, PC5 and PC10, respectively);
\item Scaling 2: we implement the PC prior with tail equivalent (in terms of probability mass) to the {half-\emph{t}} with $\gamma=10$ and $\nu=1$ (denoted as EPC) .
\end{itemize}

\subsection{Results}

Comparisons are based on the simulated rejection probabilities and expected sample sizes (ESS). For those arms where the response rate is generated under the null hypothesis, rejection probabilities indicate the type-I error rate, otherwise they refer to the power ($1$ - type II error rate). Given the presence of at least one interim analysis for each simulated trial, ESS is a performance indicator of the ethical capability of stopping the trial earlier on futility/efficacy grounds with respect to each prior.

Table~\ref{tab:operating_characteristics_37} shows operating characteristics, i.e. rejection probabilities and expected sample size, for scenarios with $N_j = 37$ under all priors. As a note to interpret the rejection probabilities in Table~\ref{tab:operating_characteristics_37}, bold numbers refer to power while non-bold numbers refer to type I error rate. The rejection probabilities for scenario 1 serve as a check that type I error rate is controlled at the desired level of $0.1$ under all priors.

Regarding how the Berry's Gamma prior (G) and the uniform (U) compare to the {half-\emph{t}}, our results are in line with those by \cite{cunanan2019variance}: G performs poorly in heterogeneous cases (scenarios 2 to 4), while giving good performance in scenario 5; U is uniformly close but inferior to the HT. 

Regarding the comparison between {half-\emph{t}} and PC priors, which is our main interest, we list below three main findings which are worthwhile to point out from looking at Table~\ref{tab:operating_characteristics_37}.

\begin{itemize}
    \item The HT prior and its associated EPC prior achieve the same performance overall. Note that the rejection probabilities (detection power) associated to these priors are basically the same in scenarios from 2 to 5. 
    \item In scenario 5, where all arms are active, PC1, PC5 and PC10 outperform HT; as expected, PC1 achieves the highest power (simulation results not shown here for sake of brevity show that using $sd<1$ will achieve at least as much power as PC1).
    \item In heterogeneous scenarios (from 2 to 4) we have mixed results. Priors PC1, PC5 and PC10 can potentially achieve more power than HT, but usually at the cost of an increased type I error rate. Note that, while in scenario 2, PC priors do not improve over the HT, in scenario 3 for some choices of $sd$ (e.g., PC1 and PC5) we have a slight increase in power, but a substantial increase in type I errors (similar behaviour can be seen in scenario 4).  
\end{itemize}

To investigate further the comparison between PC priors and {half-\emph{t}} we have also looked at their performance in smaller sample size cases $N_j = 20, 26$.  Results are reported in Appendix \ref{app:additionalres}, 
where it can be observed that reducing sample size impacts the power (i.e., rejection probabilities for data generated under the alternative hypothesis), but a suitable choice of the probability cutoff $\zeta$ still permits control of type-I error rate. The previous remarks on the performance of the different priors remain unchanged for small sample size cases too. Moreover, the tendency that we see seems to indicate that as $N$ decreases, EPC achieves higher power than HT in homogeneous cases, like in scenarios 4 and 5. 

In conclusion, consistency of our findings under different sample sizes, including very low sample size, reassures us that this simulation study can offer practical guidelines in realistic basket trials, which often times are run on low number of patients.


\begin{table}[htbp]
\centering
\caption{Rejection probabilities and expected sample size (ESS) for $N_j = 37$ and scenarios from 1 to 5 (see Table \ref{tab:scenarios_interim} for description of each scenario). Among the rejection probabilities, bold numbers refer to power ($1$ - type II error rate) while non-bold numbers refer to type I error rates.} 
\label{tab:operating_characteristics_37}
\tabcolsep=0.065cm
\begin{tabular}{c l c c c c c c c c}
\hline
& & prior  & & \multicolumn{4}{c}{rejection probabilities} & &  ESS \\
& &  & & arm 1 & arm 2 & arm 3 & arm 4 & &  \\
\hline
\parbox[c]{2mm}{\multirow{7}{*}{\rotatebox[origin=c]{90}{\textbf{scenario 1}}}} 
& & G & & 0.102 & 0.097 & 0.104 & 0.095 & & 119.9 \\
& & U & & 0.100 & 0.100 & 0.098 & 0.097 & & 126.1 \\
& & HT & & 0.103 & 0.101 & 0.097 & 0.101 & & 126.0 \\
& & PC1 & & 0.099 & 0.104 & 0.099 & 0.101 & & 122.5 \\
& & PC5 & & 0.101 & 0.104 & 0.096 & 0.097 & & 125.2 \\
& & PC10 & & 0.102 & 0.103 & 0.098 & 0.096 & & 125.7 \\
& & EPC & & 0.102 & 0.101 & 0.097 & 0.100 & & 125.9 \\
\hline
\parbox[c]{2mm}{\multirow{7}{*}{\rotatebox[origin=c]{90}{\textbf{\textbf{scenario 2}}}}} & & G & & 0.254 & 0.252 & 0.254 & \bf 0.434 & & 135.3 \\
& & U & & 0.172 & 0.166 & 0.155 & \bf 0.776 & & 130.9 \\
& & HT & & 0.179 & 0.169 & 0.159 & \bf 0.781 & & 131.0 \\
& & PC1 & & 0.206 & 0.215 & 0.205 &\bf  0.733 & & 132.8 \\
& & PC5 & & 0.184 & 0.176 & 0.172 & \bf 0.769 & & 131.6 \\
& & PC10 & & 0.182 & 0.174 & 0.167 & \bf 0.775 & & 131.2 \\
& & EPC & & 0.179 & 0.170 & 0.159 & \bf 0.781 & & 131.0 \\
\hline
\parbox[c]{2mm}{\multirow{7}{*}{\rotatebox[origin=c]{90}{\textbf{\textbf{scenario 3}}}}} & & G & & \bf 0.641 & 0.398 & 0.414 & \bf 0.641 & & 137.4 \\
& & U & & \bf 0.833 & 0.213 & 0.230 & \bf 0.851 & & 131.4 \\
& & HT & & \bf 0.836 & 0.219 & 0.234 & \bf 0.853 & & 131.6 \\
& & PC1 & & \bf 0.848 & 0.314 & 0.314 & \bf 0.859 & & 134.9 \\
& & PC5 & & \bf 0.839 & 0.236 & 0.260 & \bf 0.857 & & 132.5 \\
& & PC10 & & \bf 0.836 & 0.225 & 0.246 & \bf 0.853 & & 132.1 \\
& & EPC & & \bf 0.836 & 0.219 & 0.236 & \bf 0.852 & & 131.7 \\
\hline
\parbox[c]{2mm}{\multirow{7}{*}{\rotatebox[origin=c]{90}{\textbf{\textbf{scenario 4}}}}} & & G & & \bf 0.811 & \bf 0.799 & 0.601 & \bf 0.809 & & 128.8 \\
& & U & & \bf 0.885 & \bf 0.888 & 0.304 & \bf 0.889 & & 127.2 \\
& & HT & & \bf 0.890 & \bf 0.894 & 0.309 & \bf 0.895 & & 127.3 \\
& & PC1 & & \bf 0.921 & \bf 0.925 & 0.449 & \bf 0.920 & & 128.3 \\
& & PC5 & & \bf 0.899 & \bf 0.907 & 0.336 & \bf 0.903 & & 127.8 \\
& & PC10 & & \bf 0.892 & \bf 0.900 & 0.320 & \bf 0.899 & & 127.5 \\
& & EPC & & \bf 0.891 & \bf 0.895 & 0.309 & \bf 0.896 & & 127.4 \\
\hline
\parbox[c]{2mm}{\multirow{7}{*}{\rotatebox[origin=c]{90}{\textbf{\textbf{scenario 5}}}}} & & G & & \bf 0.931 & \bf 0.909 & \bf 0.917 & \bf 0.921 & & 110.1 \\
& & U & & \bf \bf 0.922 &\bf  0.936 & \bf 0.927 & \bf 0.928 & & 117.0 \\
& & HT & & \bf 0.925 & \bf 0.939 & \bf 0.929 & \bf 0.929 & & 117.0 \\
& & PC1 & & \bf 0.962 & \bf 0.962 & \bf 0.955 & \bf 0.964 & & 113.7 \\
& & PC5 & & \bf 0.938 & \bf 0.943 & \bf 0.936 & \bf 0.940 & & 116.3 \\
& & PC10 & & \bf 0.931 & \bf 0.942 & \bf 0.931 &\bf  0.933 & & 116.6 \\
& & EPC & & \bf 0.927 & \bf 0.941 & \bf 0.930 & \bf 0.930 & & 116.9 \\
\hline
\end{tabular}
\end{table}

\section{Discussion}
\label{sec:discussion}



Efficient strategies for phase II basket trials can be obtained by borrowing strength of information across arms via Bayesian hierarchical modelling (BHM). The user adopting such an approach will inevitably face the issue of prior choice on the arm-level variance. In this work we have reviewed the most popular priors for variance parameters in relation to their ability to handle the critical balance between pooling and locality, with special focus on the recently proposed PC prior approach. The performance attained by each prior has been studied by means of an extensive simulation study, where PC priors have been compared to the {half-\emph{t}} in several scenarios varying according to maximum sample size and level of heterogeneity between responses in each arm. 

In summary, our simulation shows that PC priors and {half-\emph{t}} priors achieve overall similar operational characteristics. In particular,  in homogeneous trials PC priors generally lead to superior designs as in larger power detection than HT; while in heterogeneous trials, PC priors have the potential to reach more power detection but at the cost of inflated type I error rates. Our simulation shows that this behaviour is consistent as $N_j$ decreases. 

We argue that PC priors offer clear advantages in terms of direct control of the balance between pooling and locality, which is crucial in basket trials. This can be handled in a practical manner by increasing/decreasing only one parameter, $\lambda$. On the contrary, tuning pooling versus locality using Gamma, Uniform and {half-\emph{t}} priors is much less intuitive and requires specification of more than one parameter. 

Finally, the scaling of the PC prior can be made intuitive to the user by linking $\lambda$ to a prior statement on the marginal standard deviation ($sd$) of the random effects: the smaller $sd$ (i.e., the larger $\lambda$) the stronger the borrowing strength of information between arms. Other intuitive approaches for choosing $\lambda$ can be based on clinical opinion on the variance associated to the odds ratios as discussed in Section \ref{sec:scaling1}.







\bibliographystyle{biorefs}
\bibliography{refs}


\newpage
\appendix

\section{Technical details about the priors presented in the paper}

\subsection{Step-by-step derivation of the PC prior for $\sigma$}
\label{app:pc}
We illustrate construction of the PC prior for the standard deviation $\sigma$ following the four principles of PC priors; for more details on the PC prior approach \citep{simpson2017penalising}.
(We want to stress that the PC prior for $\sigma$ derived below is added for illustrative purposes, it is not a novel result of this paper; it can also be derived by applying the change of variable rule to the PC prior for the precision $\tau=1/\sigma^2$, which is derived in the seminal paper by \cite{simpson2017penalising}).

\begin{itemize}
    \item \emph{Occam's razor}. Parsimony suggests that the BHM model can be seen as a flexible extension of a simpler model, denoted as base model. Let $\boldsymbol{\theta}=(\theta_1,\ldots,\theta_J)^\top$ be the vector of random effects associated to each arm, where $\boldsymbol{\theta} \sim \mathcal{N}(\boldsymbol{0},\sigma^2 \mathbf{I})$. A natural base model is the absence of random effects (i.e., $\sigma^2=0$), which implies that the response to the treatment is the same for all arms. We can formalize the flexible and base models as a $J$-dimensional Gaussian distribution denoted respectively as $\mathcal{N}_1(\boldsymbol{0},\sigma^2 \mathbf{I})$ and  $\mathcal{N}_0(\boldsymbol{0},\sigma_0^2 \mathbf{I})$, where $\sigma_0^2=0$.
    \item \emph{Model complexity}. Model complexity is measured by the distance between the (flexible) BHM and the base model, computed as $d=\sqrt{2 KLD(W||Z)}$, where $KLD$ stands for the Kullback-Leibler divergence\citep{kld-1951} from r.v. $Z$ to $W$. In our $J$-dimensional Gaussian random effects case, the $KLD$ is
    \begin{eqnarray}
    KLD(\mathcal{N}_1||\mathcal{N}_0) 
    &=& \frac{1}{2} \left[J\frac{\sigma^2}{\sigma_0^2} - J - J \ln \left( \frac{\sigma^2}{\sigma_0^2}\right)\right]
    \label{eq:kld_sigma},
    \end{eqnarray}
    which only depends on $\sigma$.
    Following \cite{simpson2017penalising} we study the behaviour of the $KLD$ as $\sigma_0^2 \to 0$. After some algebraic steps we obtain,
    \[
       KLD(\mathcal{N}_1||\mathcal{N}_0) =  \frac{J\sigma^2}{2 \sigma_0^2}\left\{ 1- \frac{\sigma_0^2}{\sigma^2} \left[ 1- \ln\left( \frac{\sigma^2}{\sigma_0^2}\right) \right] \right\}
    \longrightarrow  \frac{J\sigma^2}{2 \sigma_0^2}, \]
    for $\sigma_0 \ll \sigma$. Thus, the distance is
    \begin{equation}
        d(\sigma) = \sqrt{2 KLD(\mathcal{N}_1||\mathcal{N}_0)} = \sqrt{\frac{J\sigma^2}{ \sigma_0^2}} = \frac{\sigma}{ \sigma_0}\sqrt{J},
    \end{equation}
    which takes value in the interval $[0,\infty)$. Note, $d \equiv 0$ when the flexible and base model coincides.
    \item \emph{Constant rate penalisation}. The PC prior is a distribution on the distance $d(\sigma) = \frac{\sigma}{ \sigma_0}\sqrt{J}$, which essentially allows to penalize models deviating from the base one.  \cite{simpson2017penalising} suggests to use an exponential distribution with rate $\phi$ on $d(\sigma)$, 
    \begin{equation}
        \pi(d(\sigma)) = \phi \exp(-\phi d(\sigma))
    \end{equation}
    The exponential distribution ensures constant rate penalization. 
    Finally, the PC prior in the scale of $\sigma$ is obtained via a change of variable,
    \begin{eqnarray}
        \pi(\sigma) & = & \phi \exp(-\phi d(\sigma))\left|\frac{\partial d(\sigma)}{\partial \sigma} \right| \nonumber\\
              & = &  \frac{\phi\sqrt{J}}{\sigma_0} \exp\left(- \frac{\phi\sqrt{J}}{\sigma_0} \sigma \right) \nonumber\\
        &=& \lambda \exp(-\lambda\sigma),
    \end{eqnarray}
    where $\lambda=\frac{\phi\sqrt{J}}{\sigma_0}$.
    Note that the PC prior is an exponential prior on the distance scale, $d(\sigma)$, and it remains as such on the standard deviation scale, $\sigma$, only the rate parameter is rescaled.
\end{itemize}

\subsection{Proof of Proposition~1 in Section 4}
\label{app:proof}
\begin{proof}
Let $T$ be a centered \mbox{Student-$t$} r.v. with scale parameter $\gamma > 0$ and $\nu > 0$ degrees of freedom, having c.d.f. 
\begin{equation*}
G_T(x \mid 0, \gamma, \nu) = \frac{1}{2} \left( 1 - I\left( \frac{x^2}{x^2 + \gamma^2 \nu}; \frac{1}{2}, \frac{\nu}{2} \right) \right).
\end{equation*}
The corresponding half r.v. $|T|$ is known to have c.d.f. $G_{\mid T \mid}(x \mid 0, \gamma, \nu) = 2G_T(x \mid 0, \gamma, \nu) - 1$ for $x > 0$, and $0$ otherwise  \citep{psarakis1990folded}. By equating the survival functions of the exponential distribution with rate $\lambda$ (i.e., the PC prior)  and the
$|T|$ with scale $\gamma > 0$ and degrees of freedom $\nu > 0$, and subsequently solving for $\lambda$ (note in the paper $\lambda$ is denoted as $\lambda_{\gamma, \nu}(x)$ to stress its dependence on $x$ and the half-t parameters), we obtain:
\begin{eqnarray*}
 1-\left[1-\exp(-\lambda  x)\right] & = &  1 - G_{\mid T \mid}(x \mid 0, \gamma, \nu) \\
 \exp(-\lambda x) & = & 2 - 2 G_T(x \mid 0, \gamma, \nu) \\ 
\lambda & = & - \frac{\ln\left[ 2 - 2 G_T(x \mid 0, \gamma, \nu) \right]}{x} \\
& = & - \frac{\ln \left[ 1 - I \left( \frac{x^2}{x^2 + \gamma^2 \nu}; \frac{1}{2}, \frac{\nu}{2} \right) \right]}{x}, \quad  \quad \gamma , \nu > 0.
\end{eqnarray*}
\end{proof}

\newpage
\section{Additional results from the simulation study in Section 5}
\label{app:additionalres}

\begin{table}[htbp]
\centering
\caption{Rejection probabilities and expected sample size (ESS) for small sample sizes, $N_j=20,26$ and scenarios from 1 to 5 (see Table 2 in the paper for description of each scenario). Among the rejection probabilities, bold numbers refer to power ($1$ - type II error rate) while non-bold numbers refer to type I error rates.} 
\label{tab:operating_characteristics_small}
\tabcolsep=0.065cm
\begin{tabular}{c c c c c c c c c c c c c c c c c c c}
\hline
& & & & & \multicolumn{6}{c}{$N_j = 20$} & & \multicolumn{6}{c}{$N_j = 26$} \\
\cmidrule{6-11} \cmidrule{14-19}
& & prior  & & & \multicolumn{4}{c}{rejection probabilities} &  & ESS & & & \multicolumn{4}{c}{rejection probabilities} &  & ESS\\
\cmidrule{6-9} \cmidrule{14-17}
& &  & & & arm 1 & arm 2 & arm 3 & arm 4 & &  & & & arm 1 & arm 2 & arm 3 & arm 4 & &  \\
\hline
\parbox[c]{2mm}{\multirow{5}{*}{\rotatebox[origin=c]{90}{\textbf{scenario 1}}}} & & HT & & & 0.099 & 0.091 & 0.095 & 0.104 & & 71.0 & & & 0.096 & 0.100 & 0.098 & 0.101 & & 92.0 \\
& & PC1 & & & 0.103 & 0.098 & 0.090 & 0.090 & & 69.7 & & & 0.101 & 0.092 & 0.098 & 0.096 & & 89.6 \\
& & PC5 & & & 0.105 & 0.096 & 0.100 & 0.100 & & 71.0 & & & 0.100 & 0.099 & 0.102 & 0.099 & & 91.4 \\
& & PC10 & & & 0.103 & 0.090 & 0.096 & 0.100 & & 70.7 & & & 0.099 & 0.100 & 0.102 & 0.102 & & 91.7 \\
& & EPC & & & 0.103 & 0.092 & 0.097 & 0.103 & & 70.9 & & & 0.096 & 0.100 & 0.097 & 0.100 & & 91.9 \\
\hline
\parbox[c]{2mm}{\multirow{5}{*}{\rotatebox[origin=c]{90}{\textbf{scenario 2}}}} & & HT & & & 0.135 & 0.136 & 0.126 & \bf 0.576 & & 72.0 & & & 0.161 & 0.152 & 0.156 & \bf 0.655 & & 94.5 \\
& & PC1 & & & 0.193 & 0.185 & 0.178 & \bf 0.528 & & 73.6 & & & 0.205 & 0.190 & 0.191 &\bf  0.611 & & 95.6 \\
& & PC5 & & & 0.150 & 0.147 & 0.140 & \bf 0.570 & & 73.0 & & & 0.171 & 0.159 & 0.163 & \bf 0.644 & & 94.7 \\
& & PC10 & & & 0.137 & 0.138 & 0.132 & \bf 0.570 & & 72.0 & & & 0.166 & 0.156 & 0.158 & \bf 0.655 & & 94.5 \\
& & EPC & & & 0.138 & 0.138 & 0.130 & \bf 0.576 & & 72.0 & & & 0.161 & 0.152 & 0.157 & \bf 0.654 & & 94.5 \\
\hline
\parbox[c]{2mm}{\multirow{5}{*}{\rotatebox[origin=c]{90}{\textbf{scenario 3}}}} & & HT & & & \bf 0.658 & 0.193 & 0.196 & \bf 0.651 & & 71.9 & & &\bf  0.739 & 0.215 & 0.211 & \bf 0.740 & & 94.8 \\
& & PC1 & & & \bf 0.677 & 0.290 & 0.293 & \bf 0.672 & & 74.6 & & & \bf 0.753 & 0.294 & 0.287 & \bf 0.740 & & 96.8 \\
& & PC5 & & & \bf \bf 0.670 & 0.217 & 0.224 & \bf 0.661 & & 73.1 & & & \bf 0.746 & 0.237 & 0.232 & \bf 0.738 & & 95.3 \\
& & PC10 & & & \bf 0.660 & 0.204 & 0.206 & \bf 0.651 & & 72.0 & & & \bf 0.753 & 0.294 & 0.287 & \bf 0.740 & & 96.8  \\
& & EPC & & & \bf 0.662 & 0.201 & 0.202 & \bf 0.655 & & 71.9 & & & \bf 0.739 & 0.214 & 0.211 & \bf 0.739 & & 94.9 \\
\hline
\parbox[c]{2mm}{\multirow{5}{*}{\rotatebox[origin=c]{90}{\textbf{scenario 4}}}} & & HT & & & \bf 0.725 & \bf 0.699 & 0.260 & \bf 0.733 & & 70.6 & & & \bf 0.796 & \bf 0.780 & 0.284 & \bf 0.806 & & 92.1 \\
& & PC1 & & & \bf 0.786 & \bf 0.770 & 0.420 & \bf 0.791 & & 72.2 & & & \bf 0.851 & \bf 0.841 & 0.425 & \bf 0.860 & & 93.4 \\
& & PC5 & & & \bf 0.744 & \bf 0.714 & 0.297 & \bf 0.750 & & 71.7 & & & \bf 0.813 & \bf 0.800 & 0.322 & \bf 0.820 & & 92.6 \\
& & PC10 & & & \bf 0.730 & \bf 0.705 & 0.278 &\bf  0.740 & & 70.7 & & & \bf 0.804 & \bf  0.786 & 0.300 & \bf 0.815 & & 92.3 \\
& & EPC & & & \bf 0.727 & \bf 0.704 & 0.267 & \bf 0.735 & & 70.6 & & & \bf 0.796 & \bf 0.781 & 0.287 & \bf 0.806 & & 92.1 \\
\hline
\parbox[c]{2mm}{\multirow{5}{*}{\rotatebox[origin=c]{90}{\textbf{scenario 5}}}} & & HT & & & \bf 0.795 & \bf 0.780 & \bf 0.808 & \bf 0.787 & & 67.7 & & &\bf  0.860 & \bf \bf 0.857 & \bf 0.865 & \bf 0.866 & & 86.7 \\
& & PC1 & & & \bf 0.878 & \bf 0.881 & \bf 0.882 &\bf  0.882 & & 67.1 & & & \bf 0.923 & \bf 0.922 & \bf 0.926 & \bf 0.917 & & 85.2 \\
& & PC5 & & &\bf  0.819 &\bf  0.809 &\bf  0.836 & \bf 0.826 & & 68.0 & & &\bf  0.883 & \bf 0.883 & \bf 0.887 & \bf 0.888 & & 86.5 \\
& & PC10 & & & \bf 0.802 & \bf 0.793 & \bf 0.824 & \bf 0.800 & & 67.5 & & & \bf 0.869 & \bf 0.868 & \bf \bf 0.876 & \bf 0.877 & & 86.6 \\
& & EPC & & & \bf 0.799 & \bf 0.789 &\bf  0.816 & \bf 0.794 & & 67.6 & & &\bf  0.861 & \bf 0.860 &\bf  0.868 & \bf 0.868 & & 86.6 \\
\hline
\end{tabular}
\end{table}

\newpage
\section{R code}
\label{app:Rcode}

We illustrate the use of the R package \texttt{INLAbhmbasket} to simulate basket trials under different priors and compute the operating characteristics. In this demo we consider the three following priors: 
\begin{enumerate}
    \item PC prior with $sd=1$ (denoted as PC1);
    \item {half-\emph{t}} with $\gamma=10$ and $\nu=1$ (denoted as HT);
    \item equivalent PC prior associated to the {half-\emph{t}} with $\gamma=1$ and $\nu=1$ (denoted as EPC).
\end{enumerate}

\begin{lstlisting}
rm(list=ls())
library(INLA)
# install INLAbhmbasket using devtools
library(devtools)
install_github("massimoventrucci/INLAbhmbasket")
library(INLAbhmbasket)

## specify the study
nsim <- 500 # num sim 
m <- 4 # num arms
N <- 37 # maximum number of patients in each arm
p_null <- 0.2 # efficacy rate under H0
p_target <- 0.35 # efficacy rate under H1
ia1_fraction <- 0.4 
step <- 0.5 # define subsequent IAs' timing (as an increment of the patients enrolled at the 1st IA)

## simulate trials in scenario 1 
# (each run of sim_basket() takes approx 1h with a MacBook Pro 2,5 GHz Intel Core i7 dual-core, 16GB ram)
res.sc1.priorPC1 <- sim_basket(nsim=nsim,
                               m=m,
                               N=N,
                                 # scenario 1: p_null in all arms
                               p_true=c(p_null,p_null,p_null,p_null),  
                               p_null=p_null,
                               p_target=p_target,
                               ia1_fraction = ia1_fraction, 
                               step = step, 
                               futility_threshold = 0.05,
                               efficacy_threshold = 0.90,
                               prior='PC',
                               parameters=c(1))

res.sc1.prior.ht <- sim_basket(nsim=nsim,
                               m=m,
                               N=N,
                               p_true=c(p_null,p_null,p_null,p_null), 
                               p_null=p_null,
                               p_target=p_target,
                               ia1_fraction = ia1_fraction, 
                               step = step, 
                               futility_threshold = 0.05,
                               efficacy_threshold = 0.90,
                               prior='half-t',
                               parameters=c(10,1))

res.sc1.priorEPC <- sim_basket(nsim=nsim,
                               m=m,
                               N=N,
                               p_true=c(p_null,p_null,p_null,p_null),  
                               p_null=p_null,
                               p_target=p_target,
                               ia1_fraction = ia1_fraction, 
                               step = step, 
                               futility_threshold = 0.05,
                               efficacy_threshold = 0.90,
                               prior='PC',
                               parameters=c(22.425))


## find cutoff prob 
# this cutoff is such that type I error rate is roughly equal to alpha_target
# it has to be done in scenario 1 where H0 is true for all arms; 
# we need to find a separate cutoff prob for each prior 
p.cut.priorPC1 <- find_cutoff(alpha_target = 0.1,
                              sim.trials = res.sc1.priorPC1,
                              m=m, 
                              p_null = p_null)
p.cut.prior.ht <- find_cutoff(alpha_target = 0.1,
                              sim.trials = res.sc1.prior.ht,
                              m=m, 
                              p_null = p_null)
p.cut.priorEPC <- find_cutoff(alpha_target = 0.1,
                              sim.trials = res.sc1.priorEPC,
                              m=m, 
                              p_null = p_null)
                              
# check that cutoff gives desired alpha_target  
# (we show this only for priorPC1)
op.PC1.sc1 <- operating_char(sim.trials = res.sc1.priorPC1,
                             m=m,
                             p_null=p_null,
                             prob_cutoff= p.cut.priorPC1)
op.PC1.sc1$"rejection probabilities" # should be roughly equal to alpha_target

## now simulate trials in scenario 2, under all priors
res.sc2.priorPC1 <- sim_basket(nsim=nsim,
                               m=m,
                               N=N,
                                 # scenario 2: p_null in the first three arms, p_target in the last
                               p_true=c(p_null,p_null,p_null,p_target),  
                               p_null=p_null,
                               p_target=p_target,
                               ia1_fraction = ia1_fraction, 
                               step = step, 
                               futility_threshold = 0.05,
                               efficacy_threshold = 0.90,
                               prior='PC',
                               parameters=c(1))

res.sc2.prior.ht <- sim_basket(nsim=nsim,
                               m=m,
                               N=N,
                               p_true=c(p_null,p_null,p_null,p_target),
                               p_null=p_null,
                               p_target=p_target,
                               ia1_fraction = ia1_fraction, 
                               step = step, 
                               futility_threshold = 0.05,
                               efficacy_threshold = 0.90,
                               prior='half-t',
                               parameters=c(10,1))

res.sc2.priorEPC <- sim_basket(nsim=nsim,
                               m=m,
                               N=N,
                               p_true=c(p_null,p_null,p_null,p_target), 
                               p_null=p_null,
                               p_target=p_target,
                               ia1_fraction = ia1_fraction, 
                               step = step, 
                               futility_threshold = 0.05,
                               efficacy_threshold = 0.90,
                               prior='PC',
                               parameters=c(22.425))

# compute the operating characteristics for scenario 2 
# (type I error rate, power and expected sample size)
op.PC1.sc2 <- operating_char(sim.trials = res.sc2.priorPC1,
                             m=m,
                             p_null=p_null,
                             prob_cutoff= p.cut.priorPC1)
op.ht.sc2 <- operating_char(sim.trials = res.sc2.prior.ht,
                            m=m,
                            p_null=p_null,
                            prob_cutoff= p.cut.prior.ht)
op.EPC.sc2 <- operating_char(sim.trials = res.sc2.priorEPC,
                             m=m,
                             p_null=p_null,
                             prob_cutoff= p.cut.priorEPC)

# look at rejection probabilities 
# (type I error rate in inactive arms, power detection in active arms)
op.PC1.sc2$"rejection probabilities"
op.ht.sc2$"rejection probabilities"
op.EPC.sc2$"rejection probabilities"

# expected sample size
op.PC1.sc2$"expected sample size"
op.ht.sc2$"expected sample size"
op.EPC.sc2$"expected sample size"
\end{lstlisting}


\end{document}